\documentclass[twocolumn,aps,prl,amsmath,amssymb,superscriptaddress,tightenlines]{revtex4-2}
\usepackage{amssymb}
\usepackage{amsmath}
\usepackage{graphicx}
\usepackage{color}
\usepackage{subfigure}
\usepackage{amsfonts}
\usepackage[colorlinks,urlcolor=blue,linkcolor=blue,citecolor=blue]{hyperref}

\def\narrowtext{\par\global\columnwidth20.5pc
	\global\hsize\columnwidth\global\linewidth\columnwidth
	\global\displaywidth\columnwidth}
\begin{document}
\title{Multistability and Critical Frustrated Superradiance in Open Anisotropic Quantum Rabi Trimer}
\author{Ya-Ping Ni}
\author{Yi-Ge Wu}
\affiliation{Key Laboratory of Low-Dimensional Quantum Structures and Quantum Control of Ministry of Education, Key Laboratory for Matter Microstructure and Function of Hunan Province, Department of Physics and Synergetic Innovation Center for Quantum Effects and Applications, Hunan Normal University, Changsha 410081, China}

\author{Jie-Qiao Liao}
\email{Contact author: jqliao@hunnu.edu.cn}
\affiliation{Key Laboratory of Low-Dimensional Quantum Structures and Quantum Control of Ministry of Education, Key Laboratory for Matter Microstructure and Function of Hunan Province, Department of Physics and Synergetic Innovation Center for Quantum Effects and Applications, Hunan Normal University, Changsha 410081, China}
\affiliation{Institute of Interdisciplinary Studies, Hunan Normal University, Changsha 410081, China} 
\affiliation{Hunan Research Center of the Basic Discipline for Quantum Effects and Quantum Technologies, Hunan Normal University, Changsha 410081, China}

\author{Jin-Feng Huang}
\email{Contact author: jfhuang@hunnu.edu.cn}
\affiliation{Key Laboratory of Low-Dimensional Quantum Structures and Quantum Control of Ministry of Education, Key Laboratory for Matter Microstructure and Function of Hunan Province, Department of Physics and Synergetic Innovation Center for Quantum Effects and Applications, Hunan Normal University, Changsha 410081, China}
\affiliation{Institute of Interdisciplinary Studies, Hunan Normal University, Changsha 410081, China} 
\affiliation{Hunan Research Center of the Basic Discipline for Quantum Effects and Quantum Technologies, Hunan Normal University, Changsha 410081, China}

\begin{abstract}
We study the nonequilibrium steady states of an open anisotropic quantum Rabi trimer. Combining semiclassical and quantum mechanical treatments, we map out the system’s phase diagram and elucidate the critical phenomena induced by anisotropic light-matter coupling, cavity dissipation, and complex photon hopping, which give rise to abundant multistable phases. Counterintuitively, the frustrated superradiant phase (FSP) can be stabilized by both positive and negative photon hopping, a feature absent in closed Rabi and Dicke trimers. We further uncover two independent dissipation thresholds for the normal superradiant phase  and the FSP. Unlike isotropic open quantum Rabi trimers, both phases are suppressed once the dissipation exceeds their respective critical values. The rotating and counter-rotating interaction terms exhibit mutually reciprocal effects on the quantum phase transitions. By analyzing universal scaling behaviors with quantum fluctuations incorporated, we reveal an unprecedented finite-fluctuation critical exponent for the FSP. Moreover, the critical exponents of individual phases are coupling-strength dependent. The multistability and FSP can be tuned by anisotropy and dissipation. Our results broaden the scope of nonequilibrium quantum phase physics in few-body systems and hold promising applications for quantum state engineering and cavity/circuit quantum electrodynamics. 	
\end{abstract}
\date{\today}
\maketitle	
\narrowtext

\textit{Introduction---}~\label{introduction}
Quantum phase transitions (QPTs) represent a core paradigm for understanding emergent many-body phenomena, with promising applications in quantum computation~\cite{Zoller 2001}, quantum metrology~\cite{liangzijiliang2011,metrology2020}, and quantum information processing~\cite{liangzixinxi2008}. In particular, QPTs in quantum Rabi and Jaynes--Cummings (JC) systems---the fundamental building blocks of light--matter interacting systems---have attracted intensive research attention in recent years~\cite{QuantumFuctuations2024,M.B. Plenio2015,M.-J. Hwang2018,M.-J. Hwang2024,Experimental2024,scaling and critical2017,Y. Hu2021,X. Xie2025,Y.-Y. Zhang2022,Y.-Y. Zhang2021,Q.-T. Xie 2014}. Most existing studies focus on equilibrium phase physics in closed few-body models, while nonequilibrium QPTs in dissipative quantum systems remain poorly understood.

Cavity dissipation and anisotropic light--matter coupling profoundly reshape quantum phase behaviors, often inducing counterintuitive nonequilibrium phenomena. Meanwhile, the implementation of synthetic gauge fields in photonic cavities enables the engineering of complex phase-dependent photon hopping~\cite{Martinis2017 NP}, providing a versatile platform for exploring exotic nonequilibrium quantum phases.

In this Letter, we study an open anisotropic quantum Rabi trimer (QRT) with ring-structured complex photon tunneling. We establish its full nonequilibrium phase diagram containing seven distinct phases, including the normal phase (NP), normal superradiant phase (NSP), and frustrated superradiant phase (FSP)~\cite{MJHwang2022PRL,JC2025}, as well as abundant multistable coexisting phases. We reveal that dissipation enables a stable FSP for both positive and negative tunnelings, a feature absent in closed QRT~\cite{Y.-Y. Zhang2021} and Dicke trimers~\cite{MJHwang2022PRL}. Two independent dissipation critical thresholds are identified for the NSP and FSP, beyond which all superradiant phases are completely suppressed. Full quantum fluctuation analysis further yields anomalous finite-fluctuation critical exponents and coupling-dependent universal scaling behaviors for the FSP, which have no counterparts in conventional closed Rabi systems~\cite{QuantumFuctuations2024,scaling and critical2017}.

\textit{Model---}~\label{Model and Hamiltonian}	
We consider an open anisotropic QRT consisting of three ring-coupled cavity-QED cells, each containing a two-level atom coupled to a single-mode cavity field. The total Hamiltonian reads ($\hbar=1$)
\begin{equation} H=\sum_{n=1}^{3}H_{\text{AR},n}+\sum_{n=1}^{3}J\left(e^{i\theta}a_{n}^{\dagger}a_{n+1}+\text{H.c.}\right),\label{eq:1} \end{equation}
where $J$ and $\theta$ denote the amplitude and phase of complex intercavity photon tunneling, and the on-site anisotropic Rabi Hamiltonian is~\cite{Huang2015,Huang2017,Huang2020,X. Xie2025,Q.-T. Xie 2014}
\begin{eqnarray}
H_{\text{AR},n}&=&\frac{\omega_{0}}{2}\sigma_{n}^{z}+\omega_{c}a_{n}^{\dagger}a_{n}+ g_{\text{r}}\left(a_{n}^{\dagger}\sigma_{n}^{-}+a_{n}\sigma_{n}^{+}\right) \notag\\ 
& &+ g_{\text{cr}}\left(a_{n}^{\dagger}\sigma_{n}^{+}+a_{n}\sigma_{n}^{-}\right).\label{eq:4} 
\end{eqnarray}
Here, $a_n$ ($a_n^\dagger$) is the cavity annihilation (creation) operator, $\sigma_n^{\pm,z}$ are atomic spin operators, and $\omega_c$ and $\omega_0$ are the cavity and atomic resonant frequencies. We define the anisotropy parameter $\varepsilon=g_{\text{cr}}/g_{\text{r}}$ to characterize the relative strength of counter-rotating and rotating interactions. The system reduces to the U(1)-symmetric Jaynes--Cummings trimer for $\varepsilon=0$ and recovers the isotropic QRT at $\varepsilon=1$, while $\varepsilon\neq0$ breaks U(1) symmetry and preserves a discrete $\mathbb{Z}_2$ parity symmetry.

To describe open nonequilibrium dynamics, we adopt the Lindblad master equation accounting for cavity photon loss:
\begin{equation} \frac{d\rho}{dt}=-i[H,\rho]+\sum_{n=1}^{3}\kappa\left(2a_{n}\rho a_{n}^{\dagger}-\{a_{n}^{\dagger}a_{n},\rho\}\right),\label{eq:6} \end{equation}
where $\kappa$ is the cavity decay rate. We systematically investigate nonequilibrium QPTs via complementary semiclassical mean-field and full quantum mechanical methods.

\textit{Semiclassical Phase Diagram and Dissipative Phase Control---}
We first derive the semiclassical dynamical equations in the limit $\eta\rightarrow\infty$ for cavity and atomic mean-field observables $\alpha_{n}\equiv\langle a_{n}\rangle/\sqrt{\eta}$ and $s_{n}^{(x,y,z),\pm}\equiv\langle\sigma_{n}^{(x,y,z),\pm}\rangle$, obtaining steady-state solutions via pitchfork bifurcation analysis~\cite{Suppl}. With dimensionless rescaling, we analytically obtain the QPT critical coupling boundary:
\begin{eqnarray}
\lambda_{\text{C},\pm} & = & \sqrt{v^{-1}\left[\left(1+\varepsilon^{2}\right)u\pm\sqrt{4\varepsilon^{2}u^{2}-\bar{\kappa}^{2}v}\right]},\label{eq:13}
\end{eqnarray}
with $\lambda=g_{\text{r}}/\sqrt{\omega_{0}\omega_{c}}$, $u=1+2\bar{J}\cos\theta$,
$v=\left(1-\varepsilon^{2}\right)^{2}$, $\bar{\kappa}=\kappa/\omega_{c}$,
and $\bar{J}=J/\omega_{c}$. This criterion characterizes QPTs across anisotropy and dissipation regimes and recovers the isotropic and Jaynes--Cummings limiting results.

\begin{figure}[t]
\centering
\includegraphics[width=0.5\textwidth,scale=0.32]{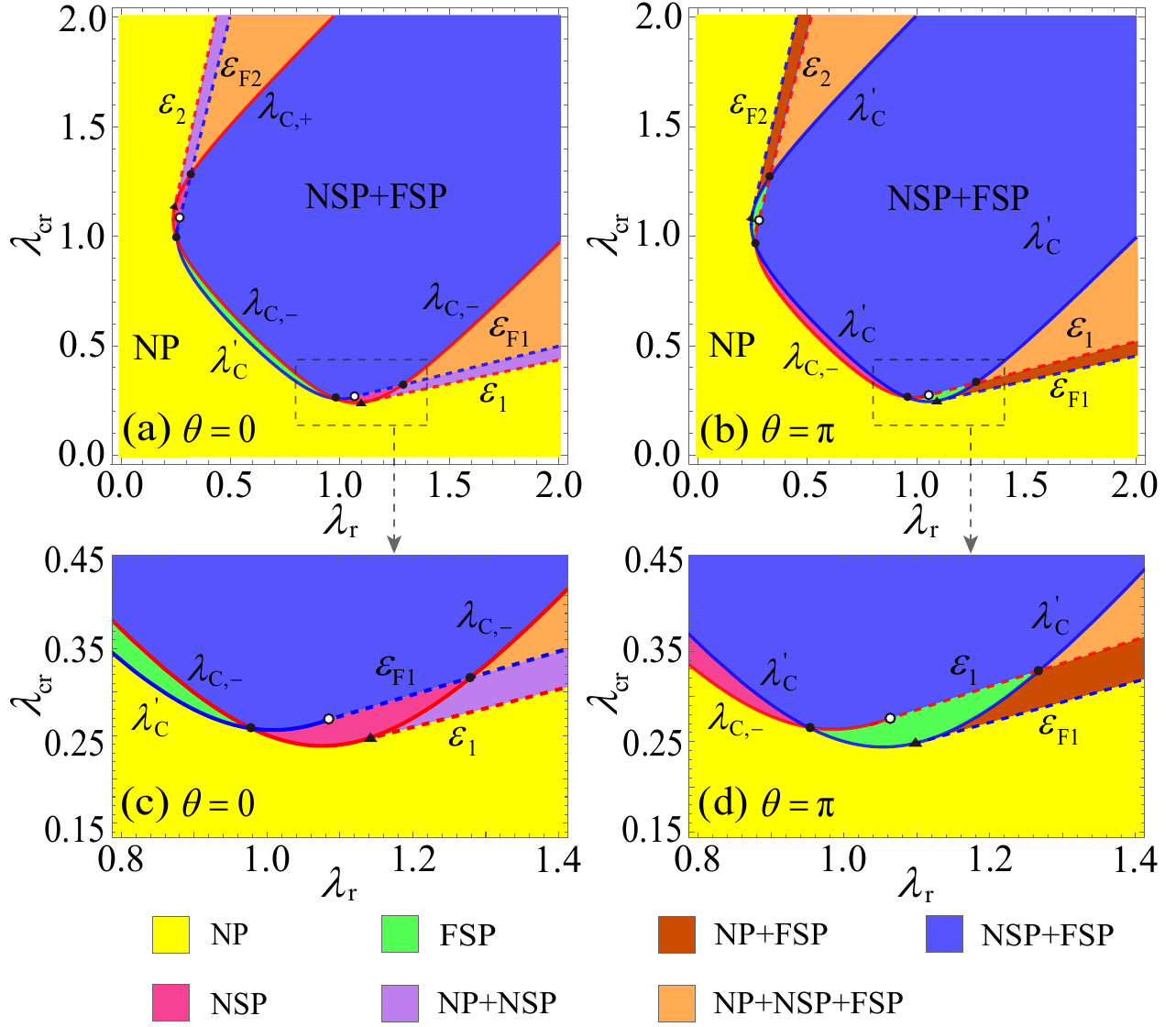}	\caption{(Color online) Phase diagram for (a,c) $\theta=0$ and (b,d) $\theta=\pi$. Panels (c) and (d) are magnified views of the black dashed-boxed regions in (a) and (b), respectively, identifying six distinct phases, multiple quadruple points, and a triple point. The red dashed lines denote the critical anisotropy parameters $\varepsilon_{1}$ and $\varepsilon_{2}$, tangent to the red solid line. The blue dashed lines $\varepsilon_{\text{F1}}$ and $\varepsilon_{\text{F2}}$, which separate distinct superradiant phases, are tangent to the blue solid line. Here we set $J/\omega_{c}=0.05$, $\kappa/\omega_{c}=0.5$.}
	\label{fig3 xiangtu}
\end{figure}

\begin{figure}[t]
	\centering
\includegraphics[width=0.48\textwidth,scale=0.22]{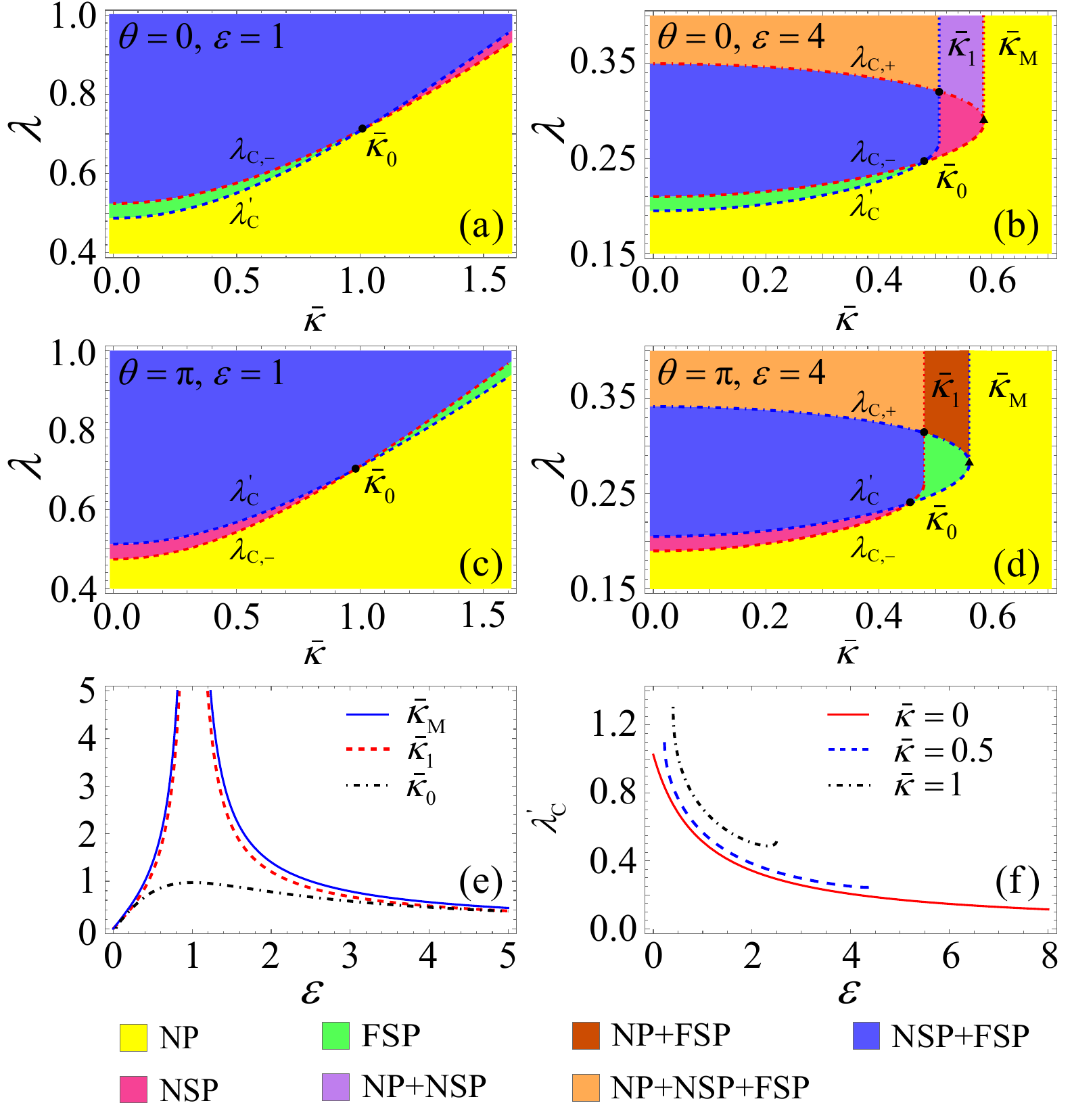}
  \caption{(Color online) Phase diagrams in the $(\bar{\kappa}, \lambda)$ plane for (a,b) $\theta=0$ and (c,d) $\theta=\pi$, with (a,c) $\varepsilon=1$ and (b,d) $\varepsilon=4$. Colored regions denote distinct quantum phases. Red dashed/dot-dashed curves denote $\lambda_{\text{C},-}$/$\lambda_{\text{C},+}$, while blue dashed curves denote $\lambda'_{C}$. The black dot marks the four-phase coexistence point at $\bar{\kappa} =\bar{\kappa}_{0}$; blue (red) dotted lines locate at critical decay rates $\bar{\kappa} =\bar{\kappa}_{\text{M}}$ ($\bar{\kappa} =\bar{\kappa}_{1}$). (e) Scaled critical decay rates $\bar{\kappa}_{j}=\kappa_{j}/\omega_{c}$ ($j=M,1,0$) vs anisotropy $\varepsilon$ at $\theta=\pi$. (f) Critical coupling $\lambda'_{C}$ vs $\varepsilon$ for $\kappa/\omega_c =0, 0.5$, and $1$ at $\theta=\pi$. We set $J/\omega_{c}=0.05$.}
\label{fig4 lambda-kappa}
\end{figure}

We construct full steady-state phase diagrams spanned by anisotropy, coupling strength, and dissipation (Figs.~\ref{fig3 xiangtu} and \ref{fig4 lambda-kappa}). The system supports nine phase regimes, including NP, NSP, FSP, and abundant multistable phases. Two anisotropy thresholds $\varepsilon_{\text{1}}$ and $\varepsilon_{\text{2}}$, together with $\varepsilon_{\text{F1}}$ and $\varepsilon_{\text{F2}}$, separate normal and superradiant domains and distinguish NSP and FSP subregimes, respectively. Superradiance is fully suppressed outside $\varepsilon\in[\varepsilon_1,\varepsilon_2]$ ($\theta=0$) and $\varepsilon\in[\varepsilon_{\text{F1}},\varepsilon_{\text{F2}}]$ ($\theta=\pi$), while FSP exclusively emerges in $\varepsilon_{\text{F1}}<\varepsilon<\varepsilon_{\text{F2}}$ above the numerically obtained threshold $\lambda_\text{C}'$. Unlike the closed Rabi and Dicke trimers where negative tunneling suppresses frustration, our dissipative system stabilizes FSP for both positive and negative $J$, revealing an open-system-specific anomalous frustration effect. High-resolution mapping resolves six fine subphases, verifying anisotropy as an effective tool to reshape quantum phase structures.

We identify two dissipation thresholds governing superradiant stability in Fig.~\ref{fig4 lambda-kappa}. In the isotropic limit $\varepsilon=1$ [Figs.~\ref{fig4 lambda-kappa}(a) and (c)], there are four phase regimes. SP and FSP persist for all dissipation values, with NP--SP boundaries strongly $\kappa$-dependent. For $\theta=0$, at low dissipation $\bar{\kappa}<\bar{\kappa}_0$, increasing coupling drives $\text{NP}\rightarrow\text{FSP}\rightarrow\text{NSP+FSP}$, while the transition sequence becomes $\text{NP}\rightarrow\text{NSP}\rightarrow\text{NSP+FSP}$ for $\bar{\kappa}>\bar{\kappa}_0$. For $\theta=\pi$, the above two transition sequences are swapped between $\bar{\kappa}<\bar{\kappa}_0$ and $\bar{\kappa}>\bar{\kappa}_0$. FSP exists for $\lambda>\lambda_\text{C}'$ both for $\theta=0$ and $\theta=\pi$. Multistable phase occurs in $\lambda>\max\left[\lambda_\text{C}',\lambda_{\text{C},-}\right]$ for all $\kappa$. A multicritical point at $\bar{\kappa}=\bar{\kappa}_0$ enables four-phase coexistence.

Enhanced anisotropy [$\varepsilon=4$, Figs.~\ref{fig4 lambda-kappa}(b) and (d)] expands multistable regimes and enhances phase complexity. Six phase regimes are supported. Two more multistable phase regimes emerge due to anisotropy. Multistable phase occurs in $\lambda>\lambda_{\text{C},-}$ with $\kappa<\kappa_1$ and $\lambda>\lambda_{\text{C},+}$ with $\kappa_1<\kappa<\kappa_\text{M}$. At $\theta=0$, it occurs in $\lambda>\lambda_{\text{C}'}$ with $\kappa<\kappa_1$ and $\lambda>\lambda_{\text{C},+}$ with $\kappa_1<\kappa<\kappa_\text{M}$. Generic anisotropic systems feature dual thresholds $\kappa_{\text{M}}$ and $\kappa_1$. Superradiance survives only for $\kappa<\kappa_{\text{M}}$ at $\theta=0$, with $\lambda>\lambda_{\text{C}'}$ ($\kappa<\kappa_0$) or $\lambda>\lambda_{\text{C},-}$ ($\kappa_0<\kappa<\kappa_\text{M}$). At $\theta=\pi$, the lower bounds for SP, $\lambda_{\text{C}'}$ and $\lambda_{\text{C},-}$, are swapped compared with the case $\theta=0$, whereas excessive dissipation locks the system in NP. For $\theta=0$, two new multistable phases $\text{NP+NSP+FSP}$ (orange) and $\text{NP+NSP}$ (purple) occur compared with $\varepsilon=1$. While for $\theta=\pi$, the new multistable phases become $\text{NP+NSP+FSP}$ (orange) and $\text{NP+FSP}$ (brown). FSP stability is tunneling-sign independent, in stark contrast to the closed-system counterparts.

Phase topology is qualitatively identical for $\theta=0$ and $\theta=\pi$, with tunneling phase only quantitatively shifting boundaries. Rotating and counterrotating interactions exhibit symmetric behavior under $\varepsilon\leftrightarrow1/\varepsilon$ [Fig.~\ref{fig4 lambda-kappa}(e)]. Superradiance vanishing at $\varepsilon=0$ confirms that weak counterrotating couplings render the superradiant phase susceptible to cavity loss.

The FSP threshold $\lambda_\text{C}'$ decreases with anisotropy but increases with cavity dissipation [Fig.~\ref{fig4 lambda-kappa}(f)], verifying dissipative suppression of the superradiant phase. Larger dissipation narrows the FSP anisotropy window, with $\varepsilon_{\text{F1}}$ rising and $\varepsilon_{\text{F2}}$ decreasing with increasing $\kappa$, establishing a robust dissipative-anisotropic phase-control mechanism.

\begin{figure}[t]
	\centering
	\includegraphics[width=0.48\textwidth]{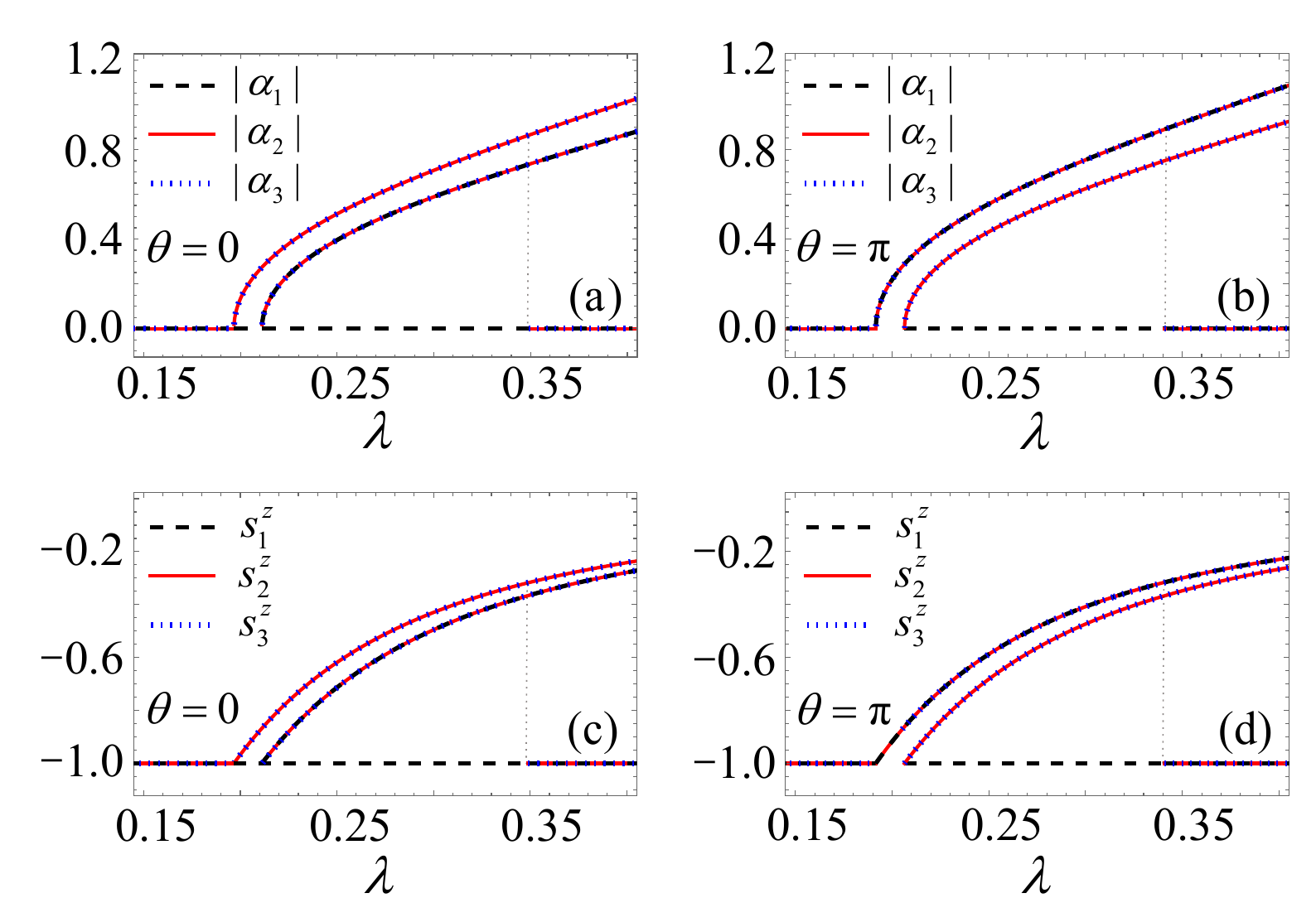}
	\caption{(Color online) Optical coherence $|\alpha_{n}|$ and spin expectation values $s_{n}^{z}$ as functions of the scaled coupling $\lambda$ at $\theta=0$ and $\theta=\pi$. Here we set $J/\omega_{c}=0.05$, $\kappa/\omega_{c}=0.1$, and $\varepsilon=4$.}
	\label{fig5 order parameter}
\end{figure}

\textit{Order parameters---}~\label{Order parameter}
To quantitatively distinguish competing quantum phases and characterize their QPT behaviors, we employ the cavity-field coherence $|\alpha_n|$ and atomic spin polarization $s_n^z$ as order parameters in the system. Here we focus on the strongly anisotropic regime $\varepsilon=4$, where rich multistable phase structures emerge, with the full order-parameter dependence on normalized coupling $\lambda$ illustrated in Fig.~\ref{fig5 order parameter}. The phase transition sequence and symmetry-breaking patterns are highly sensitive to the tunneling phase $\theta$.

For $\theta=0$, the system remains in NP with vanishing field coherence and zero spin polarization at coupling $\lambda<\lambda_\text{C}'\approx0.197$. A discontinuous jump of $|\alpha_{2,3}|$ and $s_{2,3}^z$ occurs at $\lambda=\lambda_\text{C}'$, signaling a spontaneous symmetry breaking toward FSP with pronounced site-asymmetric ordering, while the first cavity-atom unit remains unexcited, leading to frustration~\cite{MJHwang2022PRL,JC2025}. At increased coupling $\lambda_{\text{C},-}\approx0.212$, symmetric NSP $\left(\left|\alpha_{1}\right|=\left|\alpha_{2}\right|=\left|\alpha_{3}\right|,s_{1}^{z}=s_{2}^{z}=s_{3}^{z}\right)$ emerges, leading to stable NSP--FSP phase coexistence for $\lambda>0.212$. Further increasing $\lambda$ beyond $\lambda_{\text{C},+}\approx0.349$ recovers trivial NP $\left(\alpha_{l}=0,s_{l}^{z}=0,l=1,2,3\right)$, forming a robust tristable regime accommodating NP, NSP, and FSP simultaneously.

Modulating the tunneling phase to $\theta=\pi$ substantially rearranges the QPT hierarchy. The system first undergoes a NP-to-NSP symmetry-breaking transition at $\lambda_{\text{C},-}\approx0.192$, followed by a subsequent transition to NSP--FSP coexistence at $\lambda_\text{C}'\approx0.207$. In the strong-coupling regime $\lambda>\lambda_{\text{C},+}\approx0.341$, NP states coexist with superradiant solutions, reproducing the tristable phase landscape. These order-parameter trajectories perfectly corroborate the multiphase topology and tunneling-modulated transition mechanisms revealed in the $\kappa$--$\lambda$ phase diagrams, highlighting the exotic asymmetric and multistable quantum critical behaviors of the open anisotropic QRT system.

\begin{figure}[t]
	\centering
	\includegraphics[width=0.48\textwidth]{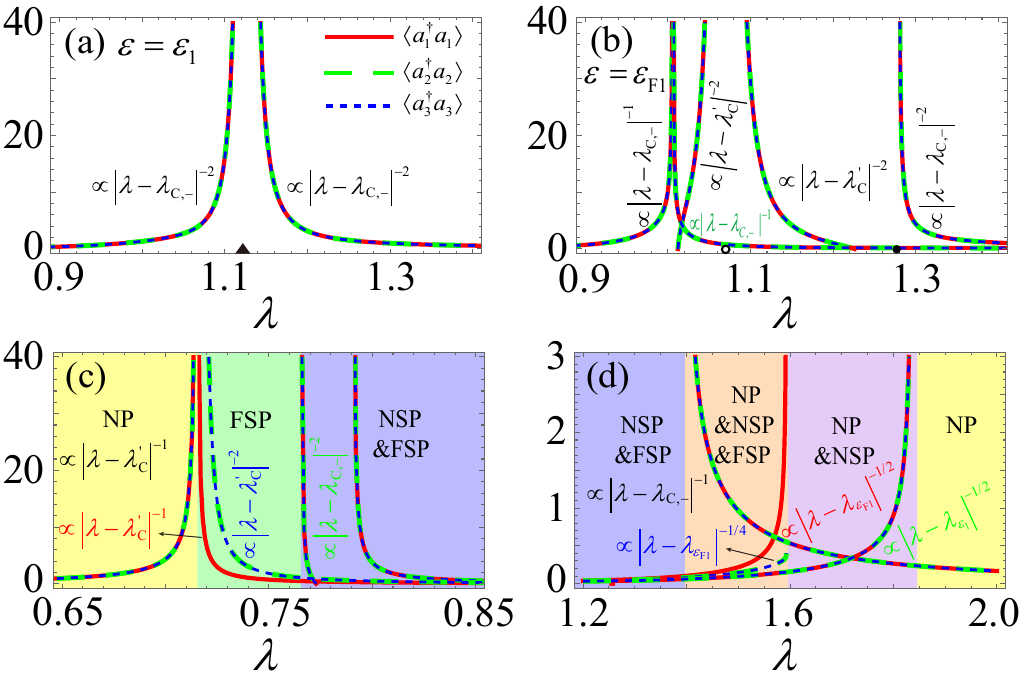}
	\caption{(Color online) Average photon number $\langle a_{n}^{\dagger}a_{n}\rangle$ as a function of the scaled coupling $\lambda$ at $\theta=0$ along (a) the red dashed line ($\varepsilon=\varepsilon_{1}$) crossing the tricritical point, (b) the blue dashed line ($\varepsilon=\varepsilon_{\text{F1}}$) crossing two quadruple points, (c) and (d) the line $\lambda_{\text{cr}}=0.4$ in Fig.~\ref{fig3 xiangtu}(a), respectively. Here we set $J/\omega_{c}=0.05$ and $\kappa/\omega_{c}=0.5$.}
	\label{fig9 NP guangzishu}
\end{figure}

\textit{Quantum fluctuation---}
To unveil the quantum critical behavior and its universality distinct from semiclassical predictions, we derive the full quantum mechanical solution by including quantum fluctuations. In the NP, the diagonalized effective Hamiltonian reads
\begin{equation}
    H_{\text{np,eff}}=\sum_{q}\zeta_{q}a_{q}^{\dagger}a_{q}+E_{g},
\end{equation}
with the quasiparticle excitation energy $\zeta_{q}=\frac{1}{2}\left[\sqrt{(\omega_{q}+\omega_{-q})^{2}-16\omega_{c}^{2}\lambda_{r}^{2}\lambda_{\text{cr}}^{2}}+\omega_{q}-\omega_{-q}\right]$ and the ground-state energy $E_{g}=\sum_{q}\frac{1}{2}(\zeta_{q}-\omega_{q})+E_{0}$. We adopt the discrete Fourier transformation $a_{n}^{\dagger}=\frac{1}{\sqrt{N}}\sum_{q}e^{inq}a_{q}^{\dagger}$ with momentum indices $q=0,\pm2\pi/3$. Solving the equations of motion for cavity field operators yields the critical rotating coupling strength
\begin{equation}
\lambda_{c\pm,q}=\sqrt{v^{-1}\left[\left(1+\varepsilon^{2}\right)u_{q}\pm\sqrt{w_{q}}\right]},
\end{equation}
where $u_{q}=1+2\bar{J}\cos q\cos\theta$ and $w_{q}=4\varepsilon^{2}u_{q}^{2}+v(i\bar{\kappa}-2\bar{J}\sin q\sin\theta)^{2}$. The counter-rotating critical coupling further satisfies $\lambda_{\text{cr}\pm,q}=\varepsilon\lambda_{c\pm,q}$. Notably, the $q=0$ quantum critical point perfectly recovers the semiclassical result $\lambda_{\text{cr}\pm,0}=\lambda_{\text{C},\pm}$. In contrast, the QPT at $q=\pm2\pi/3$ is strongly modulated by the phase parameter $\theta$. QPT survives for $\theta=n\pi$ with critical couplings consistent with semiclassical predictions, while vanishes for generic $\theta\neq n\pi$ due to complex-valued critical couplings induced by finite tunneling, anisotropy, and dissipation. The relation $\lambda_{\text{c}\pm,\pm2\pi/3}=\lambda_{\text{C}'}$ can be  approximately obtained.

For the SP, we implement cavity-field displacement via $\mathcal{D}(\alpha)=\prod_{n=1}^{N}e^{\alpha_{n}a_{n}^{\dagger}-\alpha_{n}^{*}a_{n}}$ and perform a unitary transformation to project the system onto a rotated spin-down subspace $|\tilde{\downarrow}\rangle$, yielding the effective SP Hamiltonian
\begin{eqnarray}
    \tilde{H}_{\text{sp}}^{\tilde{\downarrow}} &=& \sum_{n=1}^{3}\left[\omega_{c,n}a_{n}^{\dagger}a_{n}+J(e^{i\theta}a_{n}^{\dagger}a_{n+1}+e^{-i\theta}a_{n}^{\dagger}a_{n-1})\right] \notag\\
    & &-\sum_{n=1}^{3}\Delta_{n}^{-1}\left(O_{n}P_{n}a_{n}^{\dagger2}+Q_{n}R_{n}a_{n}^{2}\right)+E_{s}, 
\end{eqnarray}\label{eq42}
where the renormalized cavity frequency is $\omega_{c,n}=\omega_{c}-O_{n}R_{n}/\Delta_{n}-P_{n}Q_{n}/\Delta_{n}$ with the coefficients are defined in~\cite{Suppl}.

We numerically characterize quantum criticality via the site-resolved average photon number $\langle a_{n}^{\dagger}a_{n}\rangle$ in Fig.~\ref{fig9 NP guangzishu}, which exhibits universal power-law scaling $\langle a_{n}^{\dagger}a_{n}\rangle\propto|\lambda-\lambda_{c}|^{-\beta_{n}}$ near QPT points with critical exponent $\beta_{n}$. Unlike the semiclassical regime where photon occupation remains finite across phase boundaries, quantum fluctuations induce divergent photon numbers at quantum criticality. For $\varepsilon=\varepsilon_{1}$, the system features $\beta_{n}=2$ at $\lambda_{\text{C},-}=1.126$, belonging to the same universality class as the tricritical point of the anisotropic quantum Rabi model~\cite{M.-J. Hwang2024}. For $\varepsilon=\varepsilon_{\text{F1}}$, we observe distinct scaling behaviors: $\beta_{n}=1$ at standard critical points and $\beta_{n}=2$ at the tangent quantum critical point.

Systematic scans along $\lambda_{\text{cr}}=0.4$ reveal rich site-dependent critical behavior and multiple phase transitions. The system undergoes a NP-to-FSP transition at $\lambda_{\text{C}}'\approx0.716$, with uniform $\beta_{n}=1$ in the low-coupling regime $\lambda<0.716$. Strikingly, high-coupling ($\lambda>0.716$) critical exponents become site asymmetric: $\beta_{1}=1$ for the first cavity while $\beta_{2,3}=2$ for the other two cavities, exhibiting a feature distinct from single-site and closed QRT~\cite{M.-J. Hwang2024,Y.-Y. Zhang2021}.

Further increasing $\lambda$ drives the system into a multistable coexistence phase ($0.766<\lambda<1.847$), where a new NSP emerges alongside the original FSP, with uniform $\beta_{n}=1$. The photon number exhibits nonmonotonic evolution driven by atom--cavity coupling, including sharp suppression and re-divergence near successive critical points. A triple-phase coexistence regime (NP--NSP--FSP) emerges for $1.391<\lambda<1.594$, with NP branches characterized by $\beta_{n}=1$.

More intriguingly, we uncover unprecedented fractional critical exponents at higher-coupling quantum critical points. At $\lambda\approx1.594$, the first cavity obeys
\begin{eqnarray}
\langle a_{1}^{\dagger}a_{1}\rangle\propto|\lambda-\lambda_{\varepsilon_{\text{F1}}}|^{-1/2},
\end{eqnarray}
with $\beta_{1}=1/2$, while the other two cavities exhibit a novel scaling
\begin{eqnarray}
    \langle a_{2,3}^{\dagger}a_{2,3}\rangle\propto|\lambda-\lambda_{\varepsilon_{\text{F1}}}|^{-1/4},
\end{eqnarray}
with $\beta_{2,3}=1/4$, and the fluctuations remain finite. This is distinct from conventional closed-system quantum phase transitions featuring divergent quantum fluctuations at phase boundaries. This unconventional behavior arises from the joint suppression of quantum criticality by cavity dissipation and light--matter anisotropy, representing a previously unreported critical feature for QRT systems. At the final NSP--NP transition $\lambda\approx1.847$, all cavities recover $\beta_{n}=1/2$. The exotic critical exponent $\beta=1/4$ observed in our coupled cavity--atom system has no counterparts in conventional single-site quantum Rabi models, highlighting the collective quantum fluctuation effect of multi-cavity coupling.

\textit{Discussion and Conclusion.---}\label{sec:conclusion}
The anisotropic quantum Rabi model has been experimentally realized in circuit-QED platforms~\cite{Zheng2023PRL,HFan2026} and ion-trap setups~\cite{Kim2026}. Complex photon hopping with tunable phases has been demonstrated using synthetic gauge fields~\cite{Martinis2017 NP,Fan2019Sci,Baba2022SA,Oliver2024NP} and other interaction-engineering techniques~\cite{Bloch2016NC}. The quantum Rabi superradiant phase has been observed across multiple experimental platforms, including superconducting circuits~\cite{Zheng2023PRL}, ion traps~\cite{LMDuan2023NC}, and nuclear magnetic resonance systems~\cite{XhPeng2021NC}. Anisotropic couplings can be adjusted via frequency modulation~\cite{Huang2017,Huang2020,X. Xie2025} or phase-kick protocols~\cite{Huang2015}, while steady-state photon populations are measurable through cavity transmission spectroscopy. These experimental advances provide a viable route to verify our predictions. The defining signature of the FSP is its persistence under both positive and negative photon hopping. The separate dissipation thresholds for the normal and FSP phases could be resolved by sweeping cavity loss and monitoring multistable switching, offering a practical pathway for exploring nonequilibrium frustrated quantum criticality. This Letter identifies unique emergent phenomena in dissipative few-body cavity-QED systems and advances our fundamental understanding of open-system quantum criticality.

We have investigated nonequilibrium quantum phase transitions in an open anisotropic quantum Rabi trimer, clarifying the cooperative interplay among cavity dissipation, anisotropic light--matter coupling, and phase-resolved intercavity photon hopping. Combining semiclassical mean-field and fully quantum calculations, we construct the full steady-state phase diagram of this dissipative few-body system. Superradiant phases and multistability emerge only within a finite anisotropy window and are completely quenched by sufficiently strong cavity loss.
The system hosts broad multistability between superradiant and normal phases. Unlike closed Rabi and Dicke trimers, where negative photon hopping removes geometric frustration~\cite{Y.-Y. Zhang2021,MJHwang2022PRL,JC2025,H.T.Diep2005,Y.furukawa2015}, our dissipative setup stabilizes the frustrated superradiant phase (FSP) for both positive and negative photon hopping. Quantum-fluctuation analysis reveals exotic coupling-dependent universal scaling and anomalous critical exponents for the FSP, features absent in single-site~\cite{scaling and critical2017,M.-J. Hwang2024} and closed Rabi systems~\cite{QuantumFuctuations2024}. Moreover, the open system exhibits nondivergent quantum fluctuations at phase boundaries. Anisotropy and dissipation serve as versatile tuning knobs to control quantum criticality, multistability, and the FSP, enabling selective suppression or preservation of QPTs.

 \textit{Acknowledgments.} J.-F. H. is partially supported by the National Natural Science Foundation of China (Grant No. 12475016) and Young Talents of the Furong Program of Hunan
Province (Grant No. 2025RC3141). J.-Q. L. is supported in part by the National Natural Science Foundation of China (Grant  
No. 12575015).

\end{document}